\documentclass[11pt,a4paper]{article}
\usepackage{jheppub}
\usepackage{indentfirst}
\usepackage{amsmath}
\usepackage{bigstrut}
\usepackage{amssymb}
\usepackage{array}
\usepackage{multirow}
\usepackage{floatrow}
\usepackage{etoolbox}
\usepackage{graphicx}
\usepackage{hyperref}
\usepackage{amsthm}
\usepackage{slashed}
\usepackage{array}
\usepackage{epstopdf}
\usepackage{graphicx}
\usepackage{fancyhdr}
\usepackage[table,xcdraw]{xcolor}
\usepackage{float}
\usepackage{subfigure}
\usepackage[table,xcdraw]{xcolor}
\usepackage{hyperref}

\title{Probing Quartic Higgs Self-Interaction}
\author[a]{Tao Liu,}
\author[a]{Kun-Feng Lyu,}
\author[b]{Jing Ren,}
\author[c]{Hua Xing Zhu}

\affiliation[a]{Department of Physics, The Hong Kong University of Science and Technology,
Clear Water Bay, Kowloon, Hong Kong S.A.R., P.R.C.}
\affiliation[b]{Department of Physics, University of Toronto, Toronto, Ontario, Canada M5S1A7}
\affiliation[c]{Department of Physics, Zhejiang University, Hangzhou, P.R.C.}

\emailAdd{taoliu@ust.hk}
\emailAdd{klyuaa@connect.ust.hk}
\emailAdd{jren@physics.utoronto.ca}
\emailAdd{zhuhx@zju.edu.cn}

\abstract{The Higgs self-interactions play a crucial role for exploring the underlying mechanisms of electroweak symmetry breaking and the nature of the phase transition involved. In this article, we propose to probe the quartic Higgs self-interaction at lepton and hadron colliders, via the di-Higgs productions. We analyze the contributions of the quartic Higgs  coupling, including the renormalization of the cubic Higgs coupling and the modification of the $VVhh$ form factor, to the vector-boson-fusion and the vector-boson associated di-Higgs productions at one-loop level. Such an effect is independent of the choice of gauge-fixing, if the quartic Higgs coupling is decoupled from other couplings in the contexts considered. Notably, a combination of these two di-Higgs productions is important for optimizing the collider sensitivities to probe the quartic Higgs coupling. With this guideline, we explore the ILC and CLIC sensitivities, and find that the ILC has a potential to measure the quartic Higgs coupling, normalized by its SM value, with a precision of $\sim \pm 25$  (500 GeV, 4 ab$^{-1}$ + 1 TeV, 2.5 ab$^{-1}$) and $\sim \pm20$ (500 GeV, 4 ab$^{-1}$ + 1 TeV, 8 ab$^{-1}$), at $1\sigma$ C.L., after marginalizing the cubic Higgs coupling in the $\chi^2$ analysis. The dependence on the renormalization scheme of the cubic Higgs coupling is discussed.}

\begin{document}
\maketitle
\unitlength = 1mm

\section{Introduction}

The Higgs self-interaction is one of the most important targets for experimentalists to measure at colliders. In the Standard Model (SM), the Higgs potential $V_\textrm{SM}=-\mu^2 H^\dag H+\lambda(H^\dag H)^2$ is fully determined by the electroweak scale $v=246\,$GeV and the Higgs mass $m_h=125\,$GeV, with $\lambda=m_h^2/2v^2$ and $\mu^2=m_h^2/2$. The cubic and quartic Higgs couplings are then completely fixed,
\begin{eqnarray}
\lambda_{3,\textrm{SM}}=\frac{3m_h^2}{v},\quad
\lambda_{4, \textrm{SM}}=\frac{3m_h^2}{v^2}\,.
\end{eqnarray}
For many reasons new physics may enter the Higgs potential, driving the electroweak phase transition (EWPT) and yielding a deviation of the Higgs self-couplings from the SM prediction. In a general context such a deviation can be parametrized as
\begin{eqnarray}\label{eq:kappa34}
V_{\textrm{self}}
=\frac{\lambda_3}{3!}h^3+\frac{\lambda_4}{4!}h^4
\equiv\frac{1}{3!}\lambda_{3,\textrm{SM}}(1+\kappa_3)h^3+\frac{1}{4!}\lambda_{4,\textrm{SM}}(1+\kappa_4)h^4 \ , 
\end{eqnarray} 
with $\kappa_3$ and $\kappa_4$ being free parameters. Pinning down the Higgs self-couplings with precision therefore is vital for probing the underlying physics and the nature of EWPT. 

The measurements of the cubic Higgs coupling via various di-Higgs productions have been extensively studied so far. At hadron colliders, the main channels include gluon fusion production, vector-boson-fusion (VBF) production, top-pair-associated production and vector boson associated (VBA) production. At lepton colliders, the dominant channels are the $Z$ boson associated production and the VBF production. The LHC has no sensitivity to the SM cubic Higgs coupling yet.  But, the high-luminosity LHC, say, $\mathcal{L}=3$\,ab$^{-1}$@14 TeV, is expected to be able to probe it with a precision of $\sim \mathcal O(1)$ in the gluon fusion production~\cite{gf:atlas,Kim:2018uty}, with an improvement of earlier analyses (see, e.g.,~\cite{Yao:2013ika,Goertz:2013kp}). At a future 100 TeV hadron collider (for discussions at 27 TeV, see, e.g.,~\cite{Goncalves:2018qas}), the cubic Higgs coupling could be measured with a higher precision. For example, in the gluon fusion channel, the cubic Higgs coupling could be measured with a precision of  percent level~\cite{He:2015spf, Contino:2016spe}. The VBF production is found to be not quite sensitive~\cite{Dolan:2015zja, Bishara:2016kjn}. The analysis for the top-pair-associated production~\cite{Englert:2014uqa, Liu:2014rva} and VBA productions~\cite{Cao:2015oxx} at a future hadron collider are still absent. As for the lepton colliders, the International linear collider (ILC) is able to measure the cubic Higgs coupling with a precision of 27\% in the $Zhh$ production at 500\,GeV with $\mathcal{L}=4$\,ab$^{-1}$, and a precision of 14\% in the $\nu\nu hh$ production at 1\,TeV with $\mathcal{L}=2.5$\,ab$^{-1}$, respectively \cite{Asner:2013psa}.  The Compact Linear Collider (CLIC) is able to measure the cubic Higgs coupling with a precision of 54\% in the $\nu\nu hh$ channel, with $\mathcal{L}=1.5$\,ab$^{-1}$ data at 1.4 TeV, and 29\%  with $\mathcal{L}=2$\,ab$^{-1}$ data at 3 TeV~\cite{Abramowicz:2016zbo}.

To fully pin down the Higgs potential, we also need to measure the quartic Higgs coupling. The traditional wisdom for this is to measure the tri-Higgs productions. However, such  measurements are known to be difficult, even at a future 100 TeV hadron collider~\cite{Plehn:2005nk}, due to the tiny cross section of tri-Higgs production and its weak dependence on the quartic Higgs coupling. The recent studies on the tri-Higgs productions in the most promising decay channel $b\bar{b}b\bar{b}\gamma\gamma$ showed that the sensitivity to probe $\kappa_4$ in the high luminosity phase of the future hadron collider, say, $\,30\textrm{ab}^{-1}@$100 TeV, is $\sim \mathcal{O}(10)$~\cite{Papaefstathiou:2015paa}\cite{Chen:2015gva} (for studies on the tri-Higgs searches in different decay channels, see~\cite{Fuks:2015hna,Kilian:2017nio}). This motivates the proposal in this article, say, to probe the quartic Higgs coupling via its loop corrections to the di-Higgs productions. We expect a combination of the di-Higgs and tri-Higgs measurements in the future to improve the precision of measuring the quartic Higgs coupling.

For the di-Higgs productions at colliders, there are two types of one-loop effects involving the quartic Higgs coupling\footnote{Unlike other di-Higgs productions, the gluon-fusion one does not involve the quartic coupling until the two-loop level. But we will not specify this subtlety below, upon the understanding.}. Both of them are independent of the choice of gauge-fixing. The first type is the renormalization of the cubic Higgs coupling $\lambda_3$\footnote{The quartic Higgs coupling also renormalizes the Higgs mass. But, it can be fully resolved by the physical Higgs mass.}, which is universal for different di-Higgs processes. The rest of the diagrams belong to the second type. They are irreducible and finite, yielding non-trivial corrections to the form factor of the relevant vertices such as $VVhh$. The two types of diagrams are reminiscent of the self-energy and the vertex corrections induced by the cubic Higgs coupling in the single Higgs production~\cite{Degrassi:2016wml}\cite{Bizon:2016wgr}, respectively.  But, there exists a generic difference. The one-loop correction of the quartic Higgs coupling to the cubic Higgs coupling is logarithmically divergent. Its renormalization necessarily introduces a renormalization-scheme dependence on the interpretation of the experimental constraints for the cubic Higgs coupling.

A full treatment of these di-Higgs productions at one-loop level needs to embed the $\kappa$ scheme, essentially a parametrization of new physics corrections to the Higgs self-couplings, into an effective field theory (EFT) for the Higgs boson (for a review, see, e.g,~\cite{Brivio:2017vri}), and then take into account the loop effects from all relevant particles. Here the EFT could be either  the SM EFT, where new particles get decoupled at a high-energy scale, or the HEFT, which is known to describe the IR limit of some composite Higgs models (for a review, see, e.g.,~\cite{Panico:2015jxa}), dilaton constructions~\cite{Halyo:1991pc,Goldberger:2008zz}, the SM extension with a non-decoupling heavy singlet scalar~\cite{Buchalla:2016bse}, etc. In these contexts, the quartic Higgs coupling is generally decoupled from other couplings relevant to the di-Higgs productions. In the HEFT, with its potential given by $V(h)=\sum_n a_n (h/v)^n$, this feature is generic. In the SMEFT, the quartic Higgs coupling becomes decoupled as long as more than one higher dimensional operators are turned on\footnote{For discussions on the SMEFT phenomenology with $\mathcal O_6$ turned on, see, e.g.,~\cite{Barger:2003rs}.}. Interestingly, we observe that the one-loop diagrams with no quartic Higgs coupling involved (whose summation is expected to be independent of gauge-fixing, and to involve the SM couplings and $\kappa_3$ only), though interfering with the tree-level $\kappa_3$ diagrams and the one-loop $\kappa_4$ diagrams, yield a NLO impact only for both the $\kappa_3$ and $\kappa_4$ sensitivity analysis at lepton colliders after a proper renormalization for $\lambda_3$. So we will ignore such diagrams below\footnote{Though a quinary Higgs coupling may appear in the BSM physics often, it has no contributions to the di-Higgs production at one-loop level, except renormalizing the cubic Higgs coupling. In that case, the effects of the quinary Higgs coupling can be fully absorbed by the counter-term.}. The QCD loop diagrams may yield non-trivial effects for the analysis at hadron colliders. In this paper for a given di-Higgs process we assume a universal QCD $K$-factor which is independent of the corrections of the Higgs self-couplings.

The rest of the paper is organized as follows. In Sec.~\ref{sec2} we will calculate the one-loop effects of the quartic Higgs coupling in renormalizing the cubic Higgs coupling and in correcting the $VVhh$ form factor. We will also discuss how to extract the $\kappa_4$ sensitivity in a way which is less dependent on the $\lambda_3$ renormalization scheme. The numerical calculations of the VBF and VBA di-Higgs productions at both lepton and hadron colliders are presented in Sec.~\ref{sec3}. We will analyze the sensitivities of the di-Higgs probe to the quartic Higgs coupling at ILC and CLIC in Sec.~\ref{sec4}. We will conclude in Sec.~\ref{sec5}.

\section{One-loop Effects of the Quartic Higgs Coupling}
\label{sec2}

\begin{figure}[ht]
\centering
\includegraphics[width=14cm]{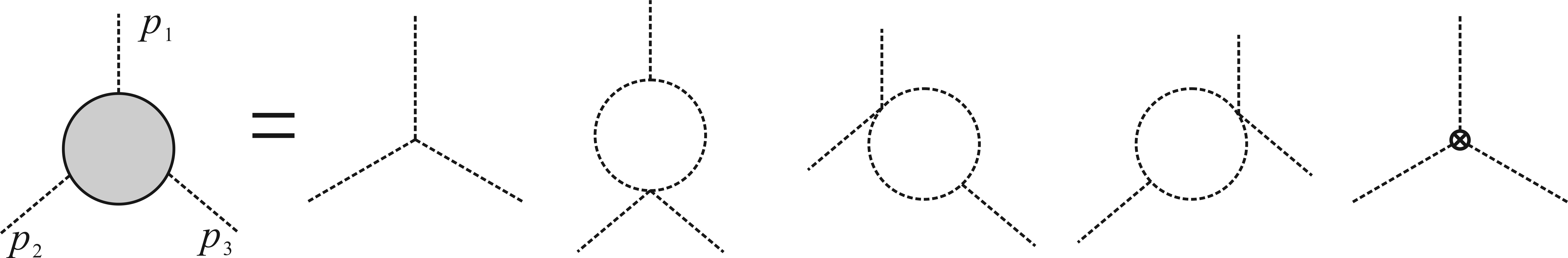}
\caption{One-loop Feynman diagrams for renormalizing the cubic Higgs coupling which are mediated by the quartic Higgs self-interaction.}
\label{fig:k3}
\end{figure}

\begin{figure}[ht]
\includegraphics[width=12cm]{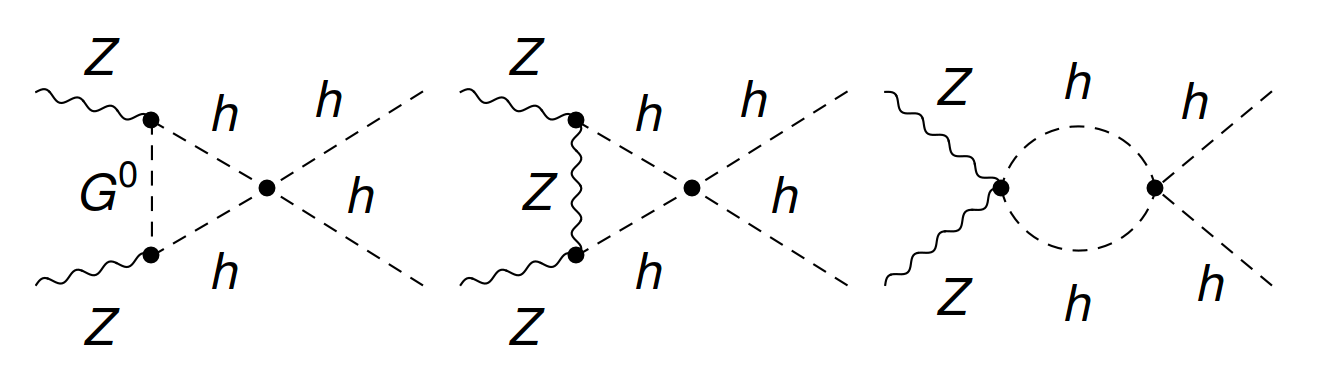}
\caption{One-loop Feynman diagrams for modifying the form factor of the quartic $VVhh$ vertex which are mediated by the quartic Higgs self-interaction. Here $G$ is Goldstone boson and we use Z boson for example.}
\label{fig:vvhh-form}
\end{figure}

\begin{figure}[h]
\centering
\includegraphics[height=3.5cm]{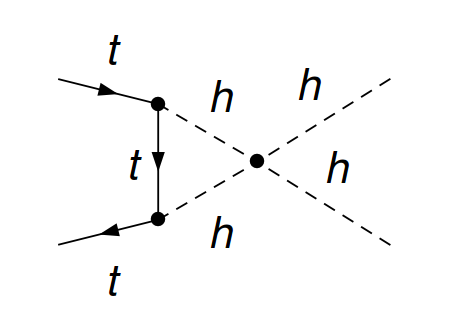}
\includegraphics[height=3.5cm]{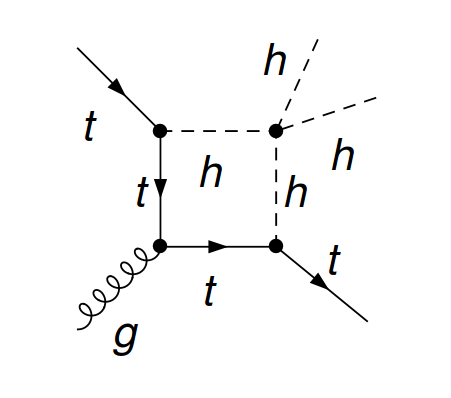}
\includegraphics[height=3.5cm]{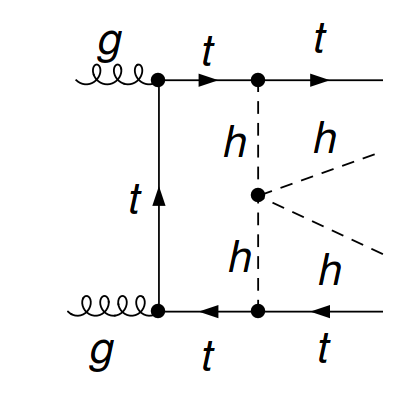}
\caption{ One-loop Feynman diagrams for modifying the form factors of the $t\bar{t}hh$, $gt\bar{t}hh$ and $ggt\bar{t}hh$ vertices which are mediated by the quartic Higgs self-interaction.}
\label{fig:ggtthh_form}
\end{figure}

As discussed above, the one-loop effects of the quartic Higgs coupling include: (1) renormalizing the cubic Higgs coupling; and (2) modifying the form factor of the relevant vertices. The relevant Feynman diagrams are shown in Fig.~\ref{fig:k3}, and  Fig.~\ref{fig:vvhh-form}-\ref{fig:ggtthh_form}, respectively. These effects are independent of the choice of gauge-fixing. For the diagrams renormalizing the cubic Higgs coupling, no Goldstone bosons or gauge bosons are involved where the gauge-fixing is applied. These diagrams will contribute to the di-Higgs productions in a universal way. For the diagrams modifying  the $VVhh$ form factor, though both the gauge bosons and Goldstone bosons are involved, their summation yields a cancellation of the gauge-dependence. These diagrams are finite and will contribute to the VBA and VBF di-Higgs productions. For the diagrams modifying the $t\bar{t}hh$, $gt\bar{t}hh$ (or  $Zt\bar{t}hh$) and $ggt\bar{t}hh$ form factors, again no Goldstone bosons or gauge bosons are involved. These diagrams are finite and will contribute to the gluon fusion and top quark associated di-Higgs productions.  Actually, as long as the quartic Higgs coupling is decoupled from other couplings in the given context, the gauge independence of its quantum corrections to the di-Higgs productions is automatically guaranteed.

Computing the diagrams in Fig.~\ref{fig:k3} with the dimensional regularization, we obtain the tri-Higgs vertex 
\begin{equation}\label{eq:renlamb3}
i\Gamma(p^2_1,p^2_2,p^2_3)  = i \lambda_{3} - i  \dfrac{\lambda_{3} \lambda_4}{32 \pi^2}\sum_{j=1}^3   \int_0^1 dx \Big(\dfrac{2}{\epsilon} - \gamma + \log(4\pi) -\log[m_h^2-x(1-x)p_j^2] \Big) + i\delta_3  \ .
\end{equation}
where $\lambda_{3}$ is the renormalized cubic Higgs coupling, and $\gamma = 0.577\ldots$ is the Euler constant. We use $\delta _3$ to denote the  counter-term schematically. This counter-term can arise from the higher dimensional operators in the SM EFT, or the $h^3$ term in the HEFT. Their coefficients then match onto the couplings between the Higgs field and the new fields in a UV complete model which have been integrated out to define the EFT. 

The renormalized cubic Higgs coupling $\lambda_3$ can be defined by properly choosing the $p_j^2$ values for $\Gamma(p^2_1,p^2_2,p^2_3)$. Since the three Higgs legs cannot be on-shell at the same time, we will consider two schemes: 
\begin{itemize}

\item Scheme 1:  set $p^2_j=0$ and define $\lambda_3\equiv\Gamma(0,0,0)$. Eq.(\ref{eq:renlamb3}) then becomes
\begin{equation}
i\Gamma(p^2_1,p^2_2,p^2_3)  = i \lambda_{3} +  i  \dfrac{\lambda_{3} \lambda_4}{32 \pi^2}\sum_{j=1}^3  \int_0^1 dx \log\left[\frac{m_h^2-x(1-x)p_j^2}{m_h^2}\right]   \ .
\end{equation}
This choice is effectively equivalent to the $\overline{\text{MS}}$ renormalization scheme with $\mu=m_h$.  

\item Scheme 2: set $p^2_{1,2}=m_h^2$, $p^2_{3}=4m_h^2$ and define $\lambda_3\equiv\Gamma(m_h^2,m_h^2,4m_h^2)$.  In any di-Higgs productions, the cubic Higgs coupling always has two on-shell Higgs legs, and the third one is characterized by the di-Higgs invariant mass $p_3^2=M_{hh}^2\geq 4m_h^2$. So we define $\lambda_3\equiv\Gamma(m_h^2,m_h^2,4m_h^2)$ and Eq. (\ref{eq:renlamb3}) becomes
\begin{equation}
i\Gamma(p^2_1,p^2_2,p^2_3)  = i \lambda_{3} +  i  \dfrac{\lambda_{3} \lambda_4}{32 \pi^2} \left(\sum_{j=1}^3  \int_0^1 dx \log\left[\frac{m_h^2-x(1-x)p_j^2}{m_h^2}\right] + 2.37\right)
\end{equation}
This choice is effectively equivalent to the $\overline{\text{MS}}$ renormalization scheme with $\mu=0.67m_h$.  
\end{itemize}

The one-loop corrections of the quartic Higgs coupling to the $VVhh$ form factor is a summation of three terms in the $R_\xi$ gauge
\begin{equation}
F[HHVV]=F_1 + F_2 +F_3 
\end{equation}
Here $F_i$ denotes the contribution of the $i$-th diagram in Fig.~\ref{fig:vvhh-form} with the momentum of the incoming gauge bosons denoted as $k_1$ and $k_2$: 
\begin{eqnarray}
F_1&=&\Big( \dfrac{i 2 m_V^2}{v} \Big)^2 \lambda_4\int \dfrac{d^4 q}{(2 \pi)^4} \Big[ \dfrac{i}{q^2 - m_V^2}\Big( -g^{\mu\nu} + \dfrac{q^\mu q^\nu}{m_V^2}\Big) - \dfrac{i}{q^2 - \xi m_V^2} \dfrac{q^\mu q^\nu}{m_V^2}  \Big]   \nonumber   \\
 &&  \dfrac{i}{(q+k_1)^2 - m_h^2} \dfrac{i}{(q+k_2)^2 - m_h^2}   \nonumber   \\
F_2&=&\Big( - \dfrac{m_V}{v} \Big)^2 \lambda_4 \int \dfrac{d^4 q}{(2 \pi)^4} \dfrac{i}{q^2 - \xi m_V^2} \dfrac{i}{(q+k_1)^2 - m_h^2} \dfrac{i (-2q-k_2)^\nu (2q+k_1)^\mu}{(q+k_2)^2 - m_h^2}  \nonumber   \\
F_3&=& \dfrac{i 2 m_V^2}{v^2}  \lambda_4\int \dfrac{d^4 q}{(2 \pi)^4} \Big[ \dfrac{i}{q^2 - m_h^2}  \dfrac{i}{(q+k_1+k_2)^2 - m_h^2} \Big]
\end{eqnarray}
After a contraction with external massless fermion current or massive gauge bosons which are on-shell, only the $q^\mu q^\nu$ term is left in $F_2$. Then the summation of $F_1$ and $F_2$ leads to a cancellation of $\xi$ dependence, as we expected. One can also check that $F[HHVV]$ is UV finite, similar to the case of $F[HVV]$ discussed in~\cite{McCullough:2013rea}. 

The calculation of  the one-loop corrections of the quartic Higgs coupling to the $t\bar{t}hh$, $gt\bar{t}hh$ and $ggt\bar{t}hh$ form factors is straightforward, based on the diagrams in Fig.~\ref{fig:ggtthh_form}. We do not show the results here, since below we will focus on the VBF and VBA di-Higgs productions.

With the renormalized cubic Higgs coupling $\lambda_3$ and the modified $VVhh$ form factor, we can parameterize the deviation of the cross section $\sigma$ from the SM prediction $\sigma_0$ in the relevant di-higgs productions as the following,  
\begin{eqnarray}
\label{eq:coef-def}
\frac{\delta\sigma}{\sigma_0}\equiv \frac{\sigma-\sigma_0}{\sigma_0}=C_{31}\kappa_3+C_{32}\kappa_3^2 + \kappa_4\left(C_{41}+ C_{42}\kappa_3+ C_{43}\kappa_3^2\right) \ ,
\end{eqnarray}
where $\kappa_3=\lambda_3 v/3m_h^2-1$. The first two terms denote the contributions from the cubic Higgs coupling only, at the leading order which arises from the tree-level. The rest arises from the interference between the $\kappa_4$ one-loop corrections and the tree-level amplitudes. We neglect the quadratic term in $\kappa_4$, given that it results from the interference between one-loop amplitudes. Then the cubic and quartic Higgs couplings can be probed by measuring the di-Higgs production cross sections at colliders. 

The interpretation of the collider sensitivities for probing $\kappa_3$ depends on the $\lambda_3$ renormalization scheme. But such a scheme dependence can be largely suppressed for $\kappa_4$, by marginalizing $\kappa_3$ in the $\chi^2$ analysis. This can be understood  in the following way. Consider $N\geq 2$ observables $\{O_i\}$ which depend on two parameters $x$ and $y$ linearly: 
\begin{eqnarray}
O_i = a_i x + b_i y \ . \label{eq:Oi}
\end{eqnarray}
The two parameters can be fit using the $\chi^2$ analysis, with 
\begin{equation}
\chi^2 = \sum_{i=1}^N \Big(\dfrac{O_i}{\sigma_i} \Big)^2 = \sum_{i=1}^N \Big(\dfrac{a_i x + b_i y}{\sigma_i} \Big)^2 \ .
\end{equation}
Here $\sigma_i$ is the measurement uncertainty of $O_i$. Then the marginalized constraint for one of the two parameters, say, $y$, can be obtained by integrating $x$ out, given by
\begin{equation}
\Delta \chi^2 = \dfrac{\det M}{M_{xx}} \Delta y^2 
= \left[\sum_{i, j=1}^N\frac{a_i^2}{\sigma_i^2}\frac{a_j^2}{\sigma_j^2}\left(\frac{b_i}{a_i}-\frac{b_j}{a_j}\right)^2 \right]\left[\sum_{i=1}^N\frac{a_k^2}{\sigma_k^2}\right]^{-1} \Delta  y^2\,.
\end{equation}
Here $M$ is the correlation matrix for $x$ and $y$. At $1\sigma$ C.L., we have $\Delta \chi^2 = 1$, which yields:
\begin{eqnarray}
\Delta y =
\sqrt{\left[\sum_{i=1}^N\frac{a_k^2}{\sigma_k^2}\right] \left[\sum_{i, j=1}^N\frac{a_i^2}{\sigma_i^2}\frac{a_j^2}{\sigma_j^2}\left(\frac{b_i}{a_i}-\frac{b_j}{a_j}\right)^2\right]^{-1} } \label{Dy}
\ . 
\end{eqnarray}

To match with the discussions on the Higgs self-couplings, we can make replacements:  $(x, y)\to(\kappa_3, \kappa_4)$ and $(a_i, b_i)\to(C_{31}^{(i)}, C_{41}^{(i)})$. At the leading order, $C_{31}^{(i)}$ is scheme-independent, but $C_{41}^{(i)}$ is not. For any given pair of observables $O_i$ and $O_j$, we can eliminate $\kappa_3$, yielding the relation
\begin{eqnarray}\label{eq:k4sub}
\frac{O_i}{C^{(i)}_{31}}-\frac{O_j}{C^{(j)}_{31}}=\left(\frac{C^{(i)}_{41}}{C^{(i)}_{31}}-\frac{C^{(j)}_{41}}{C^{(j)}_{31}}\right) \kappa_4 
\equiv \Delta C_{ij} \, \kappa_4\ .
\end{eqnarray}    
Since the left side of this equation is independent of the $\lambda_3$ renormalization scheme at the leading order, $\Delta C_{ij}$ should be nearly scheme-independent, given that $\kappa_4$ by definition is a parameter independent of $\kappa_3$ or $\lambda_3$. Then we are able to obtain $\Delta \kappa_4$ by applying Eq. (\ref{Dy}), with the scheme-dependence suppressed, if all pairs of $\{C_{41}^{(i)}, C_{41}^{(j)}\}$ are calculated with proper precisions. Note, the ``if'' condition is important for suppressing the linear-level scheme-dependence. For example, if one would combine the di-Higgs productions discussed above with the single Higgs productions in the analysis, the two-loop contributions of the quartic Higgs coupling to the latter channels need to be incorporated. The non-linear terms in Eq.~(\ref{eq:coef-def}), if turned on, may weaken this argument. But the scheme dependence introduced is of NNLO and could be further suppressed if the NNLO non-linear terms, such as the ones proportional to $\kappa_4^2$, are properly calculated.  

If there are two observables only, the formula for $\Delta \kappa_4$ is reduced to
\begin{eqnarray}
\Delta {\kappa_4} = \frac{\sqrt{(\sigma_i / C^{(i)}_{31} )^2+ (\sigma_j / C^{(j)}_{31} )^2}   }{|\Delta C_{ij}|} \ .   \label{eq:k4sub1}
\end{eqnarray}
Here $|\sigma_i / C^i_{31}|$ and $|\sigma_j / C^j_{31}|$ represent the precision of measuring $\kappa_3$ via $O^i$ and $O^j$, respectively, with $\kappa_4$ being turned off. An interesting observation is that a larger $|\Delta C_{ij}|$ tends to yield a higher precision for the $\kappa_4$ measurement. This can be the case when the two observables $O_i$, $O_j$ constrain the $\kappa_3 - \kappa_4$ plane in two clearly-separated directions. Below we will show how to optimize the measurement precision for $\kappa_4$ using this guideline.

\section{Analyses at Lepton and Hadron Colliders}
\label{sec3}

In this section we calculate the one-loop contributions of the quartic Higgs coupling in the VBF and VBA di-Higgs productions at both lepton and hadron colliders. We use FeynRule~\cite{Alloul:2013bka} to generate the model file. The cross sections are then calculated with FeynArts 3.8 and FormCalc 9.5~\cite{Hahn:2000kx} using the unitary gauge and a factorization scale of $m_h=125$ GeV, where the LoopTools~\cite{Hahn:1998yk} is linked to calculate the loop integral. The electroweak input parameters in the analysis are chosen as: $G_F=1.1663787 \times 10^{-5} \text{GeV}^{-2}$,  $m_Z=91.1876 \text{GeV}$, $m_W=80.385 \text{GeV}$ \cite{Patrignani:2016xqp}. As consistency checks, we compare the tree-level cross sections with those given by MadGraph@aMC 2.3.3~\cite{Alwall:2014hca} and CalcHEP 3.6.27~\cite{Belyaev:2012qa}. Also we have checked the values of the squared one-loop amplitudes at some given points in the phase space by comparing with the results calculated by hand. 

\subsection{Lepton Colliders}
\label{sec31}

\begin{figure}[th]
\centering
\includegraphics[width=9cm]{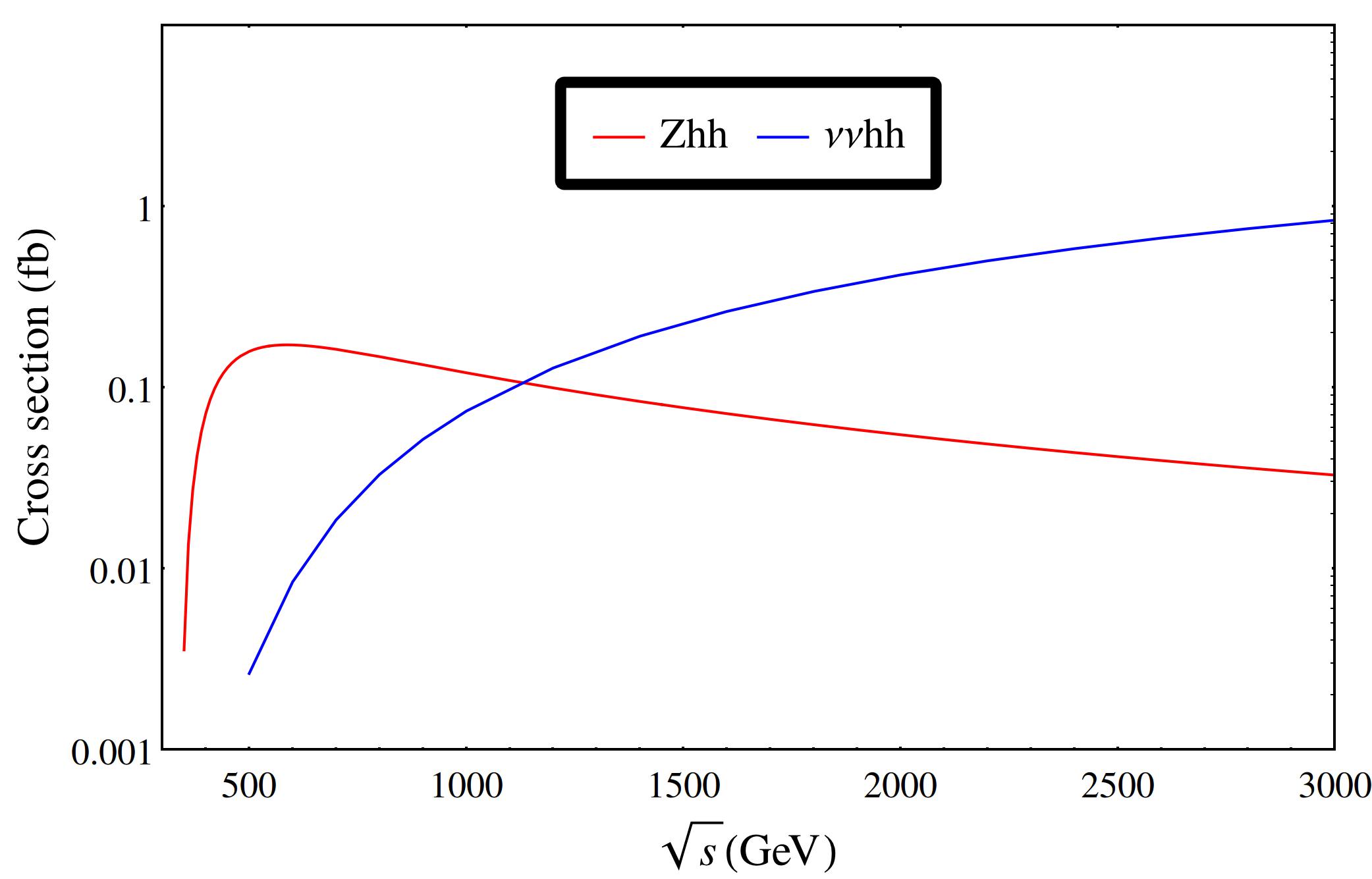}
\caption{The leading-order cross sections in the SM, as functions of the center of mass energy $\sqrt{s}$. The initial states are unpolarized.}
\label{fig:cross}
\end{figure}

At lepton colliders,  the main di-Higgs production processes include the $Z$ associated production $e^+e^- \to Zhh$ and the VBF production $e^+e^- \to \nu\nu hh$.  Though they could be kinematically turned on, the VBF production $e^+e^- \to e^+e^- hh$ and the top-pair associated production $e^+ e^- \to t \bar{t} h h$ suffer a suppression of cross section. So we will focus on the former two channels. Fig.~\ref{fig:cross} shows their leading-order cross sections in the SM, as functions of the center of mass energy $\sqrt{s}$, with an unpolarized initial state. The cross section for the $Zhh$ process reaches the peak at $\sqrt{s} \sim 500$ GeV and then slowly decreases due to a s-channel suppression. As for the VBF production of $\nu\nu h h$, due to the $t$-channel contributions mediated by the $W$ boson,  its cross section keeps growing up to a few TeV. In Table~\ref{tab:lep}, we show the leading order SM cross sections and the coefficients defined in Eq. (\ref{eq:coef-def}) for these two processes, in different collider configurations. The cubic Higgs coupling is renormalized in scheme 1. Note, the beam polarization does not modify the values of $C_{3a}$ and $C_{4b}$, but changes the total cross section only.

\begin{table}[htbp]
  \centering
  \caption{The leading-order SM cross sections and the parameterization of the $\kappa_3, \kappa_4$ contributions for the $Zhh$ and $\nu\nu hh$ di-Higgs productions at lepton colliders. Here the ILC beam is polarized as $P(e^-,e^+)=(-0.8, 0.3)$ at 500\,GeV and $P(e^-,e^+)=(-0.8, 0.2)$ at 1\,TeV.}
    \begin{tabular}{|c|c|c|c|c|c|c|c|}
    \hline
              \multicolumn{2}{|c|}{Channels}       &  $\sigma_0$ (fb) & $C_{31}$ & $C_{32}$ & $C_{41}$ & $C_{42}$ & $C_{43}$ \bigstrut\\
    \hline
    \multirow{3}[6]{*}{ILC} & $Zhh$ (500 GeV) & 0.232 & 0.564 & 0.0965 & -0.00517 & -0.00390 & -0.000810 \bigstrut\\
    \cline{2-8}          & $Zhh$ (1 TeV) & 0.166 & 0.350  & 0.0913   & -0.00271 & -0.00181 & -0.000541 \bigstrut\\
\cline{2-8}          & $\nu \nu hh$ (1 TeV) & 0.159 & -1.20  & 1.10   & -0.00327 & 0.00790 & -0.00750 \bigstrut\\
    \hline
    \multirow{3}[6]{*}{CLIC} & $Zhh$ (1.4 TeV) & 0.0833 & 0.263 & 0.0827 & -0.00186 & -0.00122 & -0.000422 \bigstrut\\
\cline{2-8}          & $\nu\nu hh$ (1.4 TeV) & 0.191 & -0.965 & 0.819 & -0.0024 & 0.00541 & -0.00505 \bigstrut\\
\cline{2-8}          & $\nu\nu hh$ (3 TeV) & 0.825 & -0.645 & 0.488 & -0.00119 & 0.00251 & -0.00247 \bigstrut\\
    \hline
    \end{tabular}%
  \label{tab:lep}%
\end{table}%

\begin{figure}[htb]
\centering
\includegraphics[height=5.5cm]{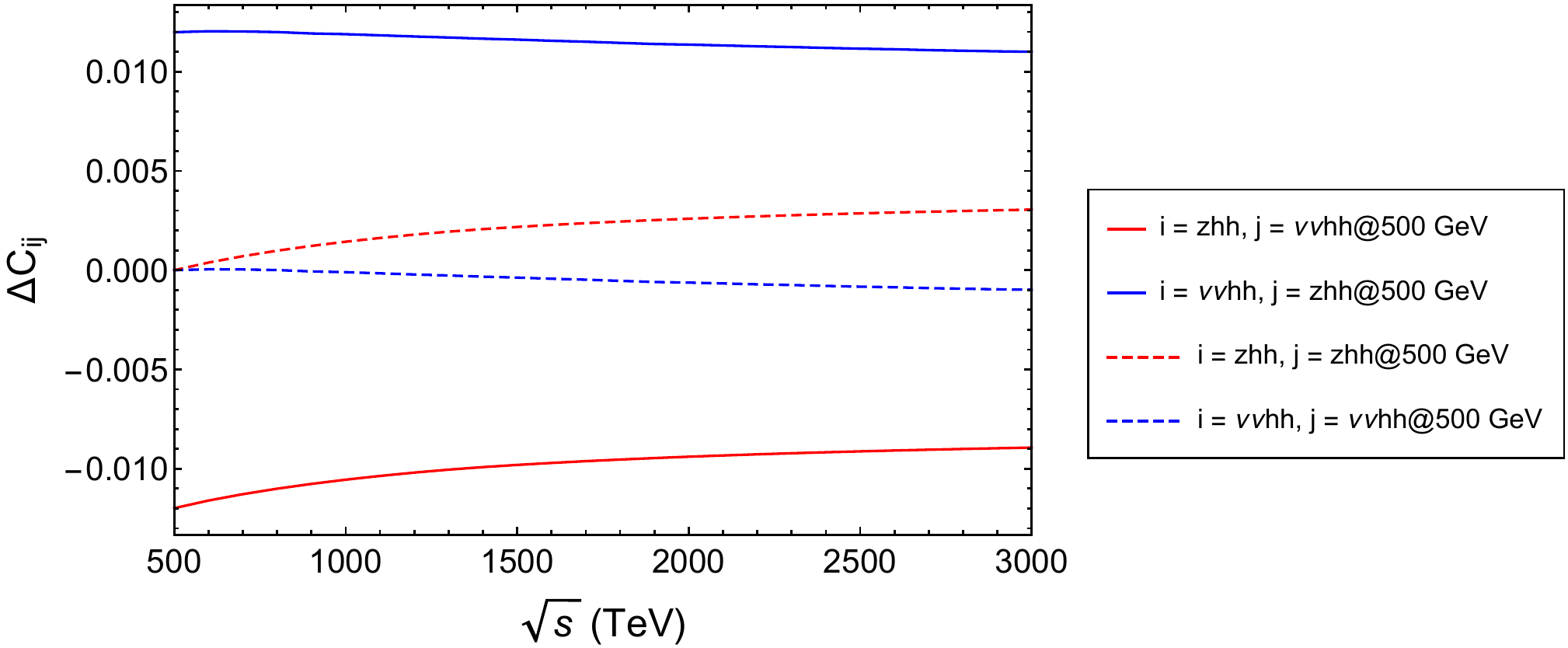}
\caption{ $\Delta C_{ij}$ for the observable pairs $\{O_i, O_j\}$ available in the $e^+e^- \to Zhh$ and $e^+e^- \to  \nu\nu hh$ channels. Here $O_j$ represents the reference observable, with $\sqrt s$ varied for $O_i$ from 500 GeV to 3 TeV.}
\label{Fig:C34Vlep}
\end{figure}

As we demonstrated in Sec.~\ref{sec2}, the $\Delta C_{ij}$ defined in Eq. (\ref{eq:k4sub}) is independent of the $\lambda_3$ renormalization scheme at the linear level. Particularly, a larger $|\Delta C_{ij}|$ tends to yield a higher precision for the $\kappa_4$ measurement, after $\kappa_3$ is marginalized. For optimizing the collider sensitivities and potentially its configuration design, therefore, it is helpful to have the information on $|\Delta C_{ij}|$ for various observable pairs available. In Fig.~\ref{Fig:C34Vlep} we show $\Delta C_{ij}$ for the observables available in  the $Zhh$ and $\nu\nu hh$ channels. The dashed and solid lines denote the cases where the two observables are from the same and different channels, respectively. The red and blue colors represent different choices for the reference observable $O_j$. Then we show the $\sqrt{s}$ dependence of $\Delta C_{ij}$ by varying $\sqrt{s}$ from 500 GeV to 3 TeV for $O_i$. Interestingly, the two observables, if arising from the $Zhh$ and $\nu\nu hh$ channels separately, result in a $|\Delta C_{ij}|$ of $\mathcal O(10^{-2})$. This is several times or even one order larger than that obtained in the complementary cases, and is not sensitive to the value of $\sqrt{s}$. Indeed, such a pair of observables have clearly-separated degenerate directions at the $\kappa_3-\kappa_4$ plane. A combination of them will be very important for optimizing the sensitivities to probe $\kappa_4$.

\subsection{Hadron Colliders}

The main di-Higgs production processes at hadron colliders include the gluon fusion production ($gg \to hh$), the top-pair-associated production ($pp \to \bar{t}thh$), the VBF production ($pp \to hhjj$) and the VBA production ($pp\to Vhh$, $V=Z, W$). For all of these processes, the cross sections increase as $\sqrt s$ increases from 14 TeV to 100 TeV. At 100 TeV, the gluon fusion cross section is around 1\,pb; the $hhjj$ and $tthh$ ones are roughly 80-90\,fb; the $Vhh$ ones are several fb~\cite{Baglio:2012np,Frederix:2014hta}. For illustration purpose, we will focus on the VBA and VBF productions.

\begin{table}[htb]
  \centering
  \caption{The leading-order SM cross sections and the parameterization of the $\kappa_3, \kappa_4$ contributions for the $Zhh/Whh$ and $jj hh$ di-Higgs productions at hadron colliders. For simplicity, we only include the contributions arising from the (anti-)up and (anti-)down quarks initiated processes. Also, we require that the $W$ boson to be electrically positive in the $Whh$ production.}
    \begin{tabular}{|c|c|c|c|c|c|c|c|}
    \hline
    \multicolumn{2}{|c|}{Channels} & $\sigma_0$ (fb) & $C_{31}$ & $C_{32}$ & $C_{41}$ & $C_{42}$ & $C_{43}$ \bigstrut\\
    \hline
    
   \multirow{3}[4]{*}{14 TeV}  &  $jjhh$  & 1.26 & -0.781 & 0.688 & -0.00233 & -0.00466 & -0.00426 \bigstrut\\
\cline{2-8}     & $Zhh$   & 0.274 & 0.496 & 0.0954 & -0.00441 & -0.00327 & -0.000738 \bigstrut\\
\cline{2-8}    & $Whh$   & 0.268 & 0.521 & 0.109 & -0.0041 & -0.00331 & -0.000807 \bigstrut\\
\hline
    \multirow{3}[4]{*}{100 TeV} & $jjhh$   & 59.3 & -0.537 & 0.411 & -0.00123 & 0.00238 & -0.00220 \bigstrut\\  
\cline{2-8}      & $Zhh$   & 2.95  & 0.454 & 0.091 & -0.00416 & -0.00293 & -0.000677 \bigstrut\\  
\cline{2-8}          & $Whh$   & 2.49  & 0.483 & 0.105 & -0.00386 & -0.003 & -0.00075 \bigstrut\\
    \hline
    \end{tabular}%
  \label{tab:hadall}%
\end{table}

Table~\ref{tab:hadall} shows the leading-order cross sections in the SM and the coefficients defined in Eq. (\ref{eq:coef-def}) for the VBA and VBF productions, at 14 and 100 TeV. Here we find the contribution from the VBF productions at hadron colliders by imposing a set of universal VBF selection cuts as the following~\cite{Bishara:2016kjn}, 
\begin{eqnarray}
p_{T,j}>25\,\textrm{GeV},\,\,
\Delta R_{jj}>4,\,\,
M_{jj}>600\,\textrm{GeV}
\end{eqnarray}
except a rapidity cut $|\eta_j| < 4.5$ at 14 TeV and $|\eta_j| < 10$ at 100 TeV. The cubic Higgs coupling is renormalized in scheme 1.

\begin{figure}[ht]
\centering
\includegraphics[width=12cm]{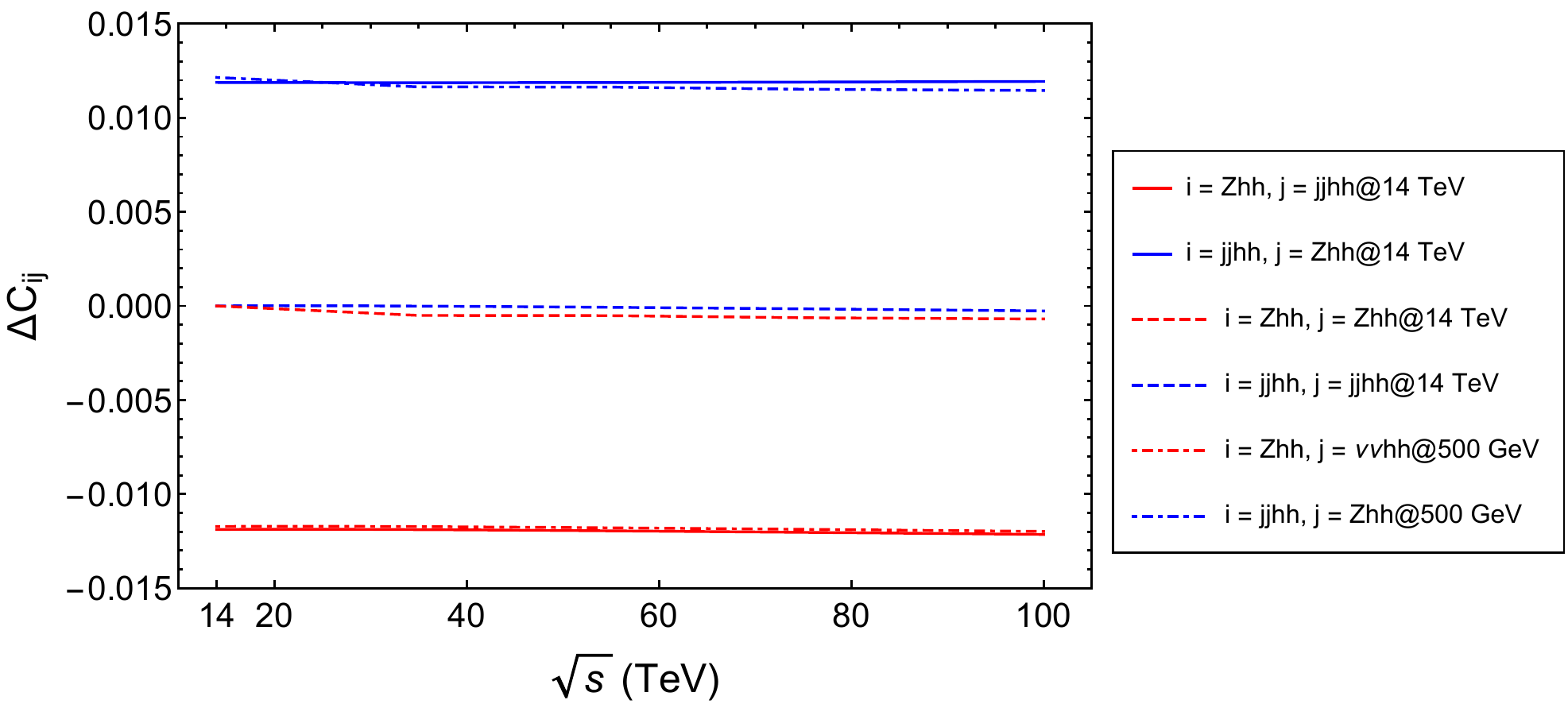}
\caption{$\Delta C_{ij}$ for the observable pairs $\{O_i, O_j\}$ available in the $pp \to Zhh$ and $pp \to  jj hh$ channels. Here $O_j$ represents the reference observable, with $\sqrt s$ varied for $O_i$ from 14 TeV to 100 TeV. }
\label{Fig:C34Vhad}
\end{figure}

Similar to the analyses at lepton colliders, the knowledge on $|\Delta C_{ij}|$ is helpful for optimizing the sensitivities at hadron collider to probe the quartic Higgs coupling. In Fig.~\ref{Fig:C34Vhad} we show $\Delta C_{ij}$ for the observable pairs which are available in the $Zhh$ and $jj hh$ channels\footnote{The $Zhh$ and $Whh$ productions share similar dependence on $\kappa_3$ and $\kappa_4$, as is indicated in Table~\ref{tab:hadall}. Considering this, we do not show the $Whh$-related curves in Fig.~\ref{Fig:C34Vhad}.}. We use the red and blue colors to denote the  $Zhh$ and the $jj hh$ as $O_i$, respectively. The lines of different styles (solid, dashed, dot-dashed) represent different reference observables $O_j$ for a given $O_i$. Then we show the $\sqrt{s}$ dependence of $\Delta C_{ij}$ by varying  $\sqrt{s}$ from 14 TeV to 100 TeV for $O_i$. The two observables, if arising from the $Zhh/Whh$ and the $jj hh$ at hadron collider separately, result in a $|\Delta C_{ij}|$ of $\mathcal O(10^{-2})$. This magnitude is several times or even one order larger than that obtained in the cases where both observables are from the $Zhh/Whh$ channels or both from the $jjhh$ channel, and is not very sensitive to the value of $\sqrt{s}$. These observations are similar to what we had at lepton colliders. So, a combination of such a pair of observables is very important for optimizing the sensitivities to probe $\kappa_4$ at hadron colliders. This conclusion can be generalized to the combination of two observables which are defined at lepton colliders and hadron colliders, separately. As is shown in Fig.~\ref{Fig:C34Vhad}, the $jj hh$ and the $Zhh$ at hadron colliders can result in a $|\Delta C_{ij}|$ of $\mathcal O(10^{-2})$ as well, by pairing with the $Zhh$ and the $\nu\nu hh$ at lepton colliders, respectively.

\section{Collider Sensitivities to the Higgs Self-couplings}
\label{sec4}

In this section we reinterpret the projected precisions of the di-Higgs measurments at lepton colliders as the sensitivities to probe both the cubic and quartic Higgs couplings.  For simplification, we assume that the Higgs self-couplings only yield negligible modifications for the signal efficiency of the SM contributions at colliders.  Then the projected precisions as summarized in Table~\ref{tab:input} can be directly applied to our analysis below, using the parameterization in Eq. (\ref{eq:coef-def}). For the convenience of discussions, we define two ILC scenarios: 
\begin{itemize}
\item ILC1 = ILC (500 GeV, 4 ab$^{-1}$ + 1 TeV, 2.5 ab$^{-1}$~\cite{Asner:2013psa}); 
\item ILC2 = ILC (500 GeV, 4 ab$^{-1}$ + 1 TeV, 8 ab$^{-1}$~\cite{Barklow:2015tja}). 
\end{itemize}

\begin{table}[ht]
  \centering
  \resizebox{\textwidth}{!}{  
    \begin{tabular}{|c|c|c|c|c|c|}  
    \hline   
     $\delta \sigma / \sigma_{\rm SM}$   & \multicolumn{3}{c|}{ILC} & \multicolumn{2}{c|}{CLIC}  \bigstrut \\                  
      \hline
   Operating Scenarios   & 500 GeV, 4 ab$^{-1}$ & 1 TeV, 2.5 ab$^{-1}$~\cite{Asner:2013psa}   & 1 TeV, 8 ab$^{-1}$~\cite{Barklow:2015tja} & 1.4 TeV, 1.5 ab$^{-1}$ & 3TeV, 3 ab$^{-1}$~\cite{CLIC:2016zwp}  \bigstrut\\    
    \hline 
    $Zhh$ & $15\%$~\cite{Asner:2013psa} &  $\textcolor[rgb]{1,0,0}{22.5\%}$~\cite{Asner:2013psa} &  $\textcolor[rgb]{1,0,0}{12.6\%}$~\cite{Asner:2013psa}  &   $\textcolor[rgb]{1,0,0}{30\%}$~\cite{Asner:2013psa} &  -  \bigstrut\\
    \hline
    $\nu\nu hh$ & - &  $16.8\%$~\cite{Asner:2013psa} & $\textcolor[rgb]{1,0,0}{9.4\%}$~\cite{Asner:2013psa}&  $44\%$~\cite{Abramowicz:2016zbo} & $16.3\%$~\cite{Abramowicz:2016zbo} \bigstrut\\
    \hline
    \end{tabular}%
  }
    \caption{Projected precision of the di-Higgs measurements at $1\sigma$ C.L., at ILC and CLIC. The numbers in red are obtained by naively rescaling the signal rates.  }
  \label{tab:input}%
\end{table}%

\begin{figure}[ht]
\centering
\includegraphics[height=11cm]{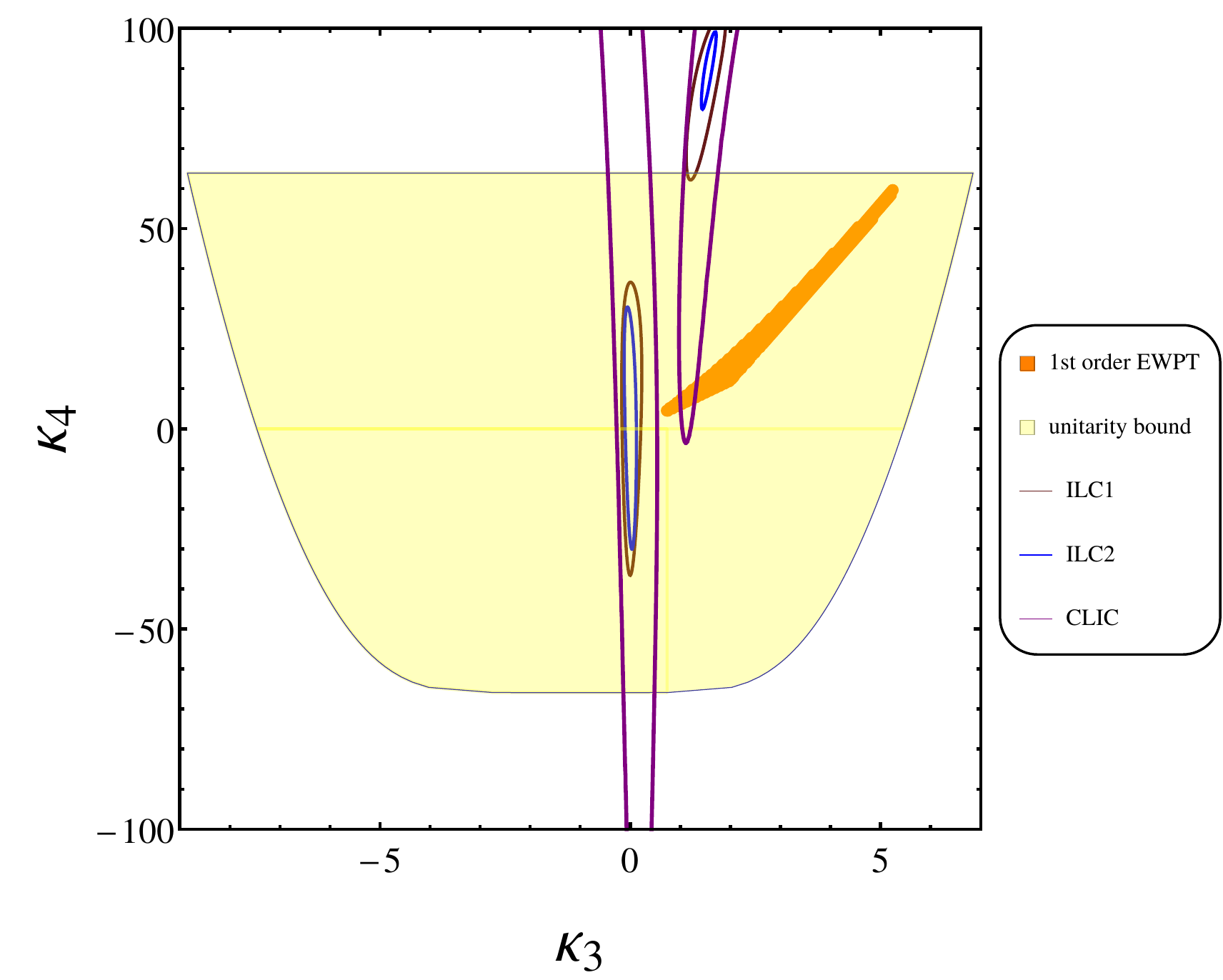}
\caption{The sensitivity contours of measuring $\kappa_3$ and $\kappa_4$ at $1\sigma$ C.L., at ILC and CLIC. The yellow region is perturbatively unitarity-safe. As a benchmark, we indicate the region in orange which is favored by first-order EWPT in the SMEFT with the $\mathcal O_6$ and $\mathcal O_8$ operators turned on (the discussions are presented in Appendix~\ref{app:1ewpt}).}
\label{Fig:2}
\end{figure}

Fig.~\ref{Fig:2} shows the sensitivity contours of measuring $\kappa_3$ and $\kappa_4$ at $1\sigma$ C.L., at ILC and CLIC. Here the cubic Higgs coupling is renormalized in scheme 1. In this figure, the yellow region is defined by the perturbative unitarity bound of the $hh \to hh$ scattering (the derivation is presented in Appendix~\ref{app:unitarity}). This unitarity requirement sets a range between $\sim \pm 65$ for $\kappa_4$, within which $\kappa_3$ is allowed to vary from $\sim -9$ to $\sim 7$.  The brown and blue circles represent the sensitivities of the ILC1 and the ILC2, respectively. In both scenarios, the ILC yields an exclusion limit for $\kappa_3$ and $\kappa_4$ well-within the perturbative regime\footnote{The circled region with $\kappa_3 \sim 1-2$ and $\kappa_4 > 60$, though not excluded by measuring di-Higgs productions at ILC, the marginalization of $\kappa_3$ yields a $\kappa_4$ completely falling outside the unitarity bound in both the ILC1 and ILC2 scenarios. Also, this region could be excluded by combining with the single Higgs productions, e.g., the $Zh$ and $\nu\nu h$ ones, at future lepton colliders~\cite{DiVita:2017vrr}. So we will not consider it here.}. This can be understood, since the ILC sensitivities benefit a lot from:
\begin{itemize}
\item the combination of the $Zhh$ and $\nu\nu h h$ observables, which are characterized by relatively large $|\Delta C_{ij}|$ values of $\mathcal O(10^{-2})$ 
\item the good precisions for measuring the $Zhh$ at 500 GeV (almost maximized cross section, high luminosity) and the $\nu\nu h h$ at 1 TeV (large cross section, high luminosity)   
\end{itemize}  
The purple circle represents the CLIC sensitivities by combining the  measurements of $Zhh$ at 1.4 GeV and $\nu\nu h h$ at 1.4 and 3 TeV. As a comparison, it is difficult for the CLIC to reach an exclusion limit for $\kappa_4$ within the perturbative regime. Its sensitivities suffer from both the suppressed $Zhh$ cross section at a higher beam energy scale, and the relatively low luminosity.

\begin{figure}[ht]
\centering
\includegraphics[width=6.3cm]{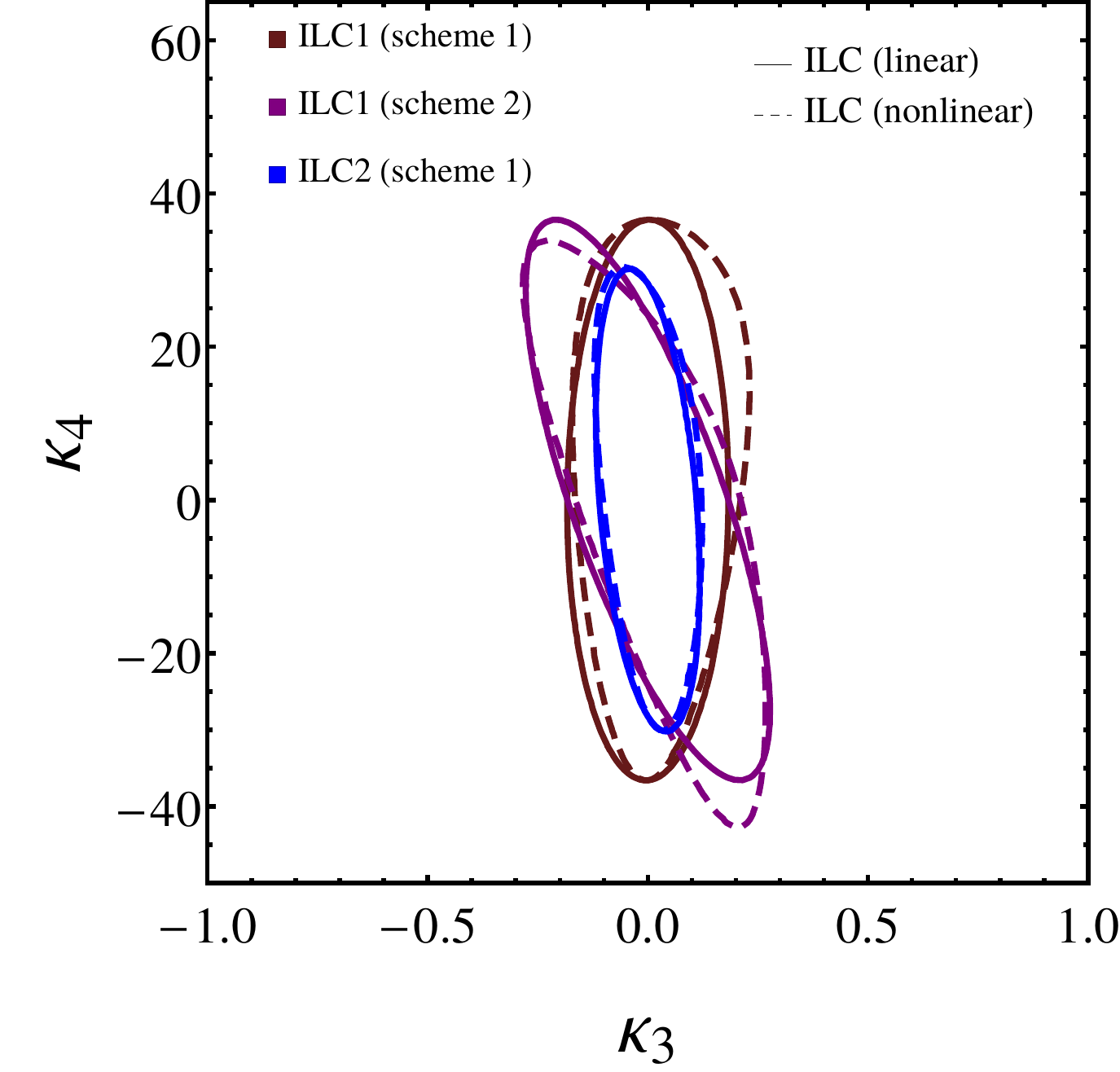}  \ \ \ \ \ \ \ \ 
\includegraphics[width=6.8cm]{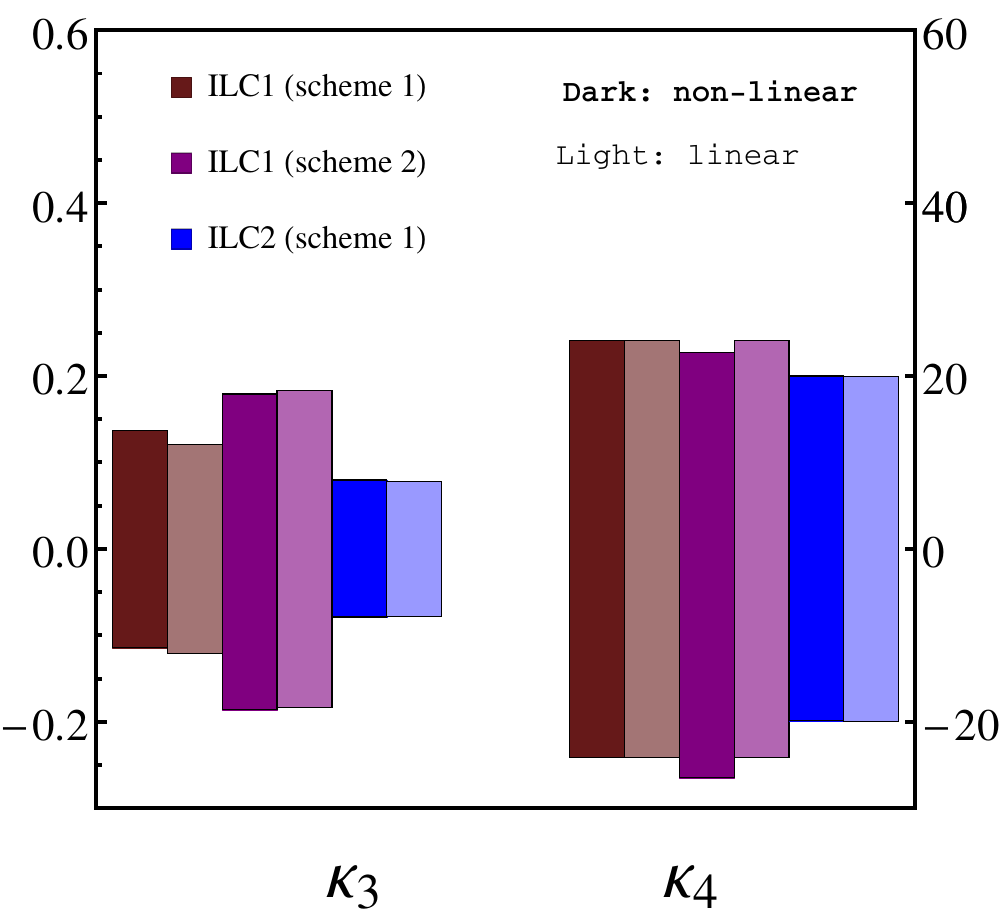}
\caption{The ILC sensitivities of measuring $\kappa_3$ and $\kappa_4$ at $1\sigma$ C.L., in different $\lambda_3$ renormalization schemes, with and without the non-linear terms in Eq. (\ref{eq:coef-def}). In the right panel, the sensitivities to probe $\kappa_3$ and $\kappa_4$ are presented with $\kappa_4$ and $\kappa_3$ marginalized, respectively. }
\label{fig:zoomin}
\end{figure}

Given its potential in probing the Higgs self-interactions, let us look into the ILC analysis and sensitivities in more detail. In Fig.~\ref{fig:zoomin} we present the ILC sensitivities in both scheme 1 and scheme 2 of the $\lambda_3$ renormalization, with and without the non-linear terms in Eq. (\ref{eq:coef-def}). At the linear level, the exclusion contours at the $\kappa_3 - \kappa_4$ plane is an ellipse with the major axis being close to the $\kappa_4$ direction. As is indicated in the left panel, the change from scheme 1 to scheme 2 yields a counter-clockwise rotation for the exclusion contours. The non-linear terms  deform these ellipses. Compared to scheme 2, the ellipse orientation in scheme 1 restricts $\kappa_3$ to be small and makes the nonlinear effects less important.  In the right panel, the sensitivities to probe $\kappa_3$ and $\kappa_4$ are shown by marginalizing  $\kappa_4$ and $\kappa_3$ respectively in the $\chi^2$ fit. Unlike the $\kappa_3$ sensitivities,  the $\kappa_4$ ones are nearly independent of the $\lambda_3$ renormalization scheme at the linear level, as we advertised in section~\ref{sec2}. The scheme-dependence is mainly  introduced via the non-linear terms in Eq. (\ref{eq:coef-def}) in this context. With these effects, the allowed ranges for $\kappa_4$ are varied by several percents between scheme 1 and scheme 2. In all, the ILC has a potential to probe $|\kappa_4|$ as small as $\sim 25$ in the ILC1 scenario and $\sim 20$ in the ILC2 scenario, respectively, at $1\sigma$ C.L., in the $\lambda_3$ renormalization scheme that we choose. Such a sensitivity is comparable to the one which could be achieved by measuring the tri-Higgs production at high-luminosity future hadron collider, say, 30 ab$^{-1}$@100 TeV~\cite{Papaefstathiou:2015paa}\cite{Chen:2015gva}.



\section{Conclusion and Outlook}
\label{sec5}

The Higgs self-interactions play a crucial role for exploring the underlying mechanisms of electroweak symmetry breaking and the nature of the phase transition involved. Motivated by this, we proposed to probe the quartic Higgs self-interaction at lepton and hadron colliders, via the di-Higgs productions. We analyzed the corrections of the quartic Higgs  coupling to the VBF and VBA di-Higgs productions at the one-loop level. Such an effect is independent of the gauge fixing, if the quartic Higgs coupling is decoupled from other couplings in the given context. In the calculations we ignored the one-loop diagrams with no quartic Higgs coupling involved. These diagrams  yield a NLO impact only for the sensitivity analysis of $\kappa_3$ and $\kappa_4$ at lepton colliders, after a proper renormalization for $\lambda_3$. One notable observation in the analysis is that the observables from the VBF and VBA di-Higgs productions probe the $\kappa_3 - \kappa_4$ plane in two clearly-separated  directions, at both lepton and hadron colliders. A combination of these two channels therefore is important for optimizing the collider sensitivities. With this guideline, we analyzed the ILC and CLIC sensitivities. We are able to extract the nearly renormalization scheme independent sensitivity on $\kappa_4$, at least at the linear level, by marginalizing the cubic Higgs coupling in the $\chi^2$ analysis. We found that the ILC has a potential to measure the quartic Higgs coupling, normalized by its SM value, with a marginalized precision of $\sim \pm 25$ in the ILC1 scenario and $\sim \pm 20$ in the ILC2 scenario, respectively, at $1\sigma$ C.L..

The collider sensitivities could be further improved by utilizing the di-Higgs invariant mass distribution of the di-Higgs events. In the analysis pursued, we have assumed that new physics does not significantly modify the kinematics of the SM di-Higgs events. To look into this further, we show the SM cross sections and the values of $C_{3a}$ and $C_{4b}$ in the di-Higgs invariant mass bins of $e^-e^+ \to Zhh$ and $e^-e^+ \to \nu\nu hh$ at ILC, in Table~\ref{tab:zhh500} and Table~\ref{tab:vvhh1000} of Appendix~\ref{app:Cvalues}, respectively. It is easy to see, though the $\frac{C_{41}}{C_{31}}$ defined in the $Zhh$ channel is not very sensitive to the $m_{hh}$ values, a relatively small $m_{hh}$ value yields a more negative $\frac{C_{41}}{C_{31}}$ in the $\nu\nu hh$ channel and hence a larger $|\Delta C_{ij} |$ between the two channels. Additionally, both channels become more sensitive to $\kappa_3$ in the low $m_{hh}$ region, with a larger $|C_{31}|$ value. According to Eq. (\ref{eq:k4sub}), therefore, the collider sensitivities could be further improved by requiring relatively small $m_{hh}$ for the di-Higgs events. Furthermore, if Eq. (\ref{eq:k4sub1}) is applied to the pair of observables $Zhh$ at 500 GeV and $\nu\nu hh$ at 1 TeV, we can check
\begin{eqnarray}
 \left(\frac{\sigma_{Zhh}}{C_{31}^{Zhh}} \right )^2  \gg  \left(\frac{\sigma_{\nu\nu hh}}{C_{31}^{\nu\nu hh}} \right )^2  \ .
\end{eqnarray}
Thus, by improving the measurement precision for the $Zhh$ at 500 GeV, if sizably, the sensitivities for probing $\kappa_4$ could be significantly improved.

We need to keep in mind, that the di-Higgs productions could be contaminated by some other new physics, via, e.g., the wave function renormalization of gauge bosons or Higgs boson, the definition shift of the EW parameters, or the introduction of new vertices. Here we have turned off all of these effects and simply assumed that they can be constrained sufficiently well for our purpose, by the electroweak and Higgs precision measurements at future colliders (for recent studies, see, e.g.,~\cite{Ellis:2015sca,Durieux:2017rsg,Barklow:2017suo,Chiu:2017yrx}). 

Given the significance of measuring the Higgs self-interactions in particle physics, it is worthwhile to pursue a more systematic and complete analysis on its collider sensitivities. We can extend the analysis from lepton colliders to hadron colliders, particularly to the next-generation hadron colliders. More di-Higgs production channels can be taken into account, such as the gluon fusion and top-quark-associated processes, in that case. The leading-order effects of the quartic Higgs coupling appear at two- and one-loop level, respectively. We may also incorporate the tri-Higgs productions at both lepton colliders and hadron colliders in the analysis. The observables arising from these channels could be characterized by a $|\Delta C_{ij}|$ of $\mathcal O(10^{-2})$ as well, and further improve the marginalized precision of $\kappa_4$. Additionally, the quartic Higgs coupling contributes to the single Higgs productions (e.g., $Zh$ and $\nu\nu h$) at two-loop level, which in turn may facilitate the probe for the quartic Higgs coupling. To end the discussion, we would stress again that to probe $\kappa_4$ by combining the di-Higgs productions and other Higgs channels, the $C_{41}/C_{31}$ for both need to be calculated with proper precisions, to suppress the scheme-dependence of the $\lambda_3$ renormalization at least at the linear level. We leave a full study on these to the future work. 

{\bf [Note added]} While this paper was in finalization, the paper~\cite{Maltoni:2018ttu} appeared, which partially overlaps with this one in analyzing the one-loop corrections of the quartic Higgs coupling to the $Zhh$ and $\nu\nu hh$ productions at lepton colliders. But, our work is different from the paper~\cite{Maltoni:2018ttu} in the following aspects: (1) we developed a general guideline for optimizing the collider sensitivities of probing the quartic Higgs coupling, based on Eq. (\ref{eq:k4sub}); (2) we analyzed the one-loop corrections of the quartic Higgs coupling to the $Zhh/Whh$ and $jj hh$ productions at hadron colliders as well; (3) we presented the ILC sensitivities for probing $\kappa_4$ by marginalizing $\kappa_3$ in the $\chi^2$ analysis, which is nearly independent of the renormalization scheme of the cubic Higgs coupling, at least at the linear level.

\begin{acknowledgments}
We would like to thank Tao Han, Sunghoon Jung, Zhen Liu and Lian-Tao Wang for useful discussions. K.F Lyu would thank Xiaozhou Li, Yu Zhang and Shaoming Wang for discussions on the tool of loop integration.  T. Liu is supported in part by the General Research Fund (GRF) under Grant No. 16312716 and in part by the GRF under Grant No. 16302117, both of which are issued by the Research Grants Council of Hong Kong S.A.R.. J. Ren is supported in part by the Natural Sciences and Engineering Research Council of Canada.
\end{acknowledgments}


\appendix

\section{Perturbative Unitarity Bound}
\label{app:unitarity}

To have the perturbative calculation still reliable, Higgs self-couplings need to satisfy the perturbative unitarity bound. The scattering amplitude for $h h \rightarrow h h$ at tree level is
\begin{equation}
\mathcal{M}(E,\theta) = -\lambda_3^2 \left(\dfrac{1}{s-m_h^2} + \dfrac{1}{t-m_h^2} + \dfrac{1}{u-m_h^2}\right)-\lambda_4
\end{equation}
where the Mandelstam variables $s=E^2$, $t= -(E^2 - 4 m_h^2) \sin^2 \theta/2$, $u=- (E^2 - 4 m_h^2) \cos^2 \theta/2$ in the center of mass frame. The partial wave amplitudes are then computed as~\cite{DiLuzio:2016sur,DiLuzio:2017tfn} 
\begin{eqnarray}
a_\ell (E)=\frac{1}{2}\frac{1}{32\pi E^2}\beta(E^2,m_h^2,m_h^2)\int^{1}_{-1}\textrm{d} \cos\theta P_\ell(\cos\theta)\mathcal{M}(E,\theta)
\end{eqnarray}
where the additional factor of $1/2$ comes from normalization of the symmetric initial, final states and the kinematic factor $\beta(x,y,z)=(x^2+y^2+z^2-2xy-2yz-2xz)^{1/2}$. For the s-wave, $\ell=0$, we find  
\begin{equation}
a_0(E) = -\dfrac{1}{32\pi} \sqrt{\frac{E^2-4 m_h^2}{E^2}} \left[ \lambda_3 \left( \dfrac{1}{E^2 - m_h^2}-\dfrac{2}{E^2-4m_h^2} \log \frac{E^2-3 m_h^2}{m_h^2} \right) + \lambda_4 \right]
\end{equation}
The s-wave unitarity condition requires: $| \textrm{Re}\, a_0(E) | < 1/2$.
In the high energy limit $E^2 \gg m_h^2$, $\lambda_4$ contributes at the leading order. Thus we can obtain $|\lambda_4 | < 16 \pi$, namely 
\begin{equation}\label{eq:k4UB}
|1 + \kappa_4 | < \dfrac{16 \pi v^2}{3 m_h^2} = 65
\end{equation}
$\lambda_3$ starts to dominate at the low energy and the amplitude reaches a peak at some scale. Assuming the peak amplitude satisfies the s-wave unitarity condition, we can find the range of $\kappa_3$ for a given $\kappa_4$ satisfying Eq. (\ref{eq:k4UB}).

\section{First-Order Electroweak Phase Transition: a Benchmark}
\label{app:1ewpt}


The nature of EWPT could have a strong correlation with the Higgs potential at zero temperature. For illustrating the collider capability in probing the EWPT nature, we analyze a simplified model in the SMEFT~\cite{Huang:2015tdv} (we will tolerate the potential uncertainties caused by such a simplified treatment~\cite{Chala:2018ari}; for discussions in more general contexts, see, e.g.,~\cite{Jain:2017sqm})
\begin{eqnarray}
V_\textrm{SMEFT}(T) = (-\mu^2+a_0 T^2) H^\dagger H +\lambda(H^\dagger H)^2 + \dfrac{c_6}{ \Lambda^2} (H^\dagger H)^3 + \dfrac{c_8}{\Lambda^4} (H^\dagger H)^4\,
\end{eqnarray}
as the benchmark. Here the temperature-dependent term results from an expansion of thermal mass for the SM particles. $a_0 \sim 3$ is defined by the SM physics. The first-order EWPT requires the coexistence of two degenerate vacua, characterized by $v=0$ and $v=v_c\neq 0$ at the critical temperature $T_c$, with the following condition satisfied:
\begin{eqnarray}
(v_T-v)^2(3c_6v^2+4c_8v^4+2c_8v^2v_T^2)+2M_h^2(v_T^2-v^2)+4a_0T^2v^2=0\,.
\end{eqnarray}
We then scan over $\{c_6,c_8\}$, to extract out the $\{\kappa_3, \kappa_4\}$ region where a first-order EWPT is favored, using the relation 
\begin{eqnarray}
\lambda_3 &=& \dfrac{3 m_h^2}{v} (1 + \dfrac{2 c_6 v^4}{m_h^2 \Lambda^2}+\dfrac{4 c_8 v^6}{m_h^2 \Lambda^4}) \nonumber \\
\lambda_4 &=& \dfrac{3 m_h^2}{v^2} (1 + \dfrac{12 c_6 v^4}{m_h^2 \Lambda^2}+\dfrac{32 c_8 v^6}{m_h^2 \Lambda^4}) \ .
\end{eqnarray}
The favored region is marked in orange in Fig.~\ref{Fig:2}. In the case with $c_8 =0$, the orange region is reduced to the bottom boundary, which is consistent with the results obtained in~\cite{Huang:2015tdv}, where only the $\mathcal O_6$ operator is turned on.

\section{$C_{3a}$ and $C_{4b}$ for Di-Higgs Productions at ILC}
\label{app:Cvalues}

\begin{table}[htb]
  \centering
    \begin{tabular}{|c|c|c|c|c|c|c|}
    \hline
          & $\sigma_0$ (fb) & $C_{31}$ & $C_{32}$ & $ C_{41}$ & $C_{42}$ & $C_{43}$ \bigstrut\\
    \hline
    $m_{hh}$ (GeV) & 0.232 &  0.564 &  0.0965  & -0.00517 & -0.0039  & -0.00081 \bigstrut\\
    \hline
    (250,300)   & 0.0647  & 0.862 & 0.195  & -0.00799 & -0.00771  & -0.00192 \bigstrut\\
    \hline
    (300,350)   & 0.0845 & 0.567  & 0.086 & -0.00516 & -0.00351  & -0.00061 \bigstrut\\
    \hline
    (350,410)   & 0.0826 & 0.328 & 0.03 & -0.00297  & -0.00131 &  -0.00014 \bigstrut\\
    \hline
    \end{tabular}%
  \caption{The SM cross sections and the parameterization of the $\kappa_3, \kappa_4$ contributions in the di-Higgs invariant mass bins of $e^-e^+ \to Zhh$ at ILC 500 GeV. Here the ILC beam is polarized as $P(e^-,e^+)=(-0.8, 0.3)$. Note, the $C_{3a}$ and $C_{4b}$ values are independent of the beam polarization.}
  \label{tab:zhh500}%
\end{table}%

\begin{table}[h]
  \centering
    \begin{tabular}{|c|c|c|c|c|c|c|}
    \hline
          & $\sigma_0$ (fb) & $C_{31}$ & $C_{32}$ & $ C_{41}$ & $C_{42}$ & $C_{43}$ \bigstrut\\
    \hline
    $m_{hh}$ (GeV) & 0.159 & -1.20  & 1.10   & -0.00327 & 0.0079 & -0.0075 \bigstrut\\
    \hline
    (250,350)   & 0.0458 & -1.96 & 2.62  & -0.00642 & 0.0189 & -0.0226 \bigstrut\\
    \hline
    (350,450)   & 0.0540 & -1.20  & 0.741 & -0.00321 & 0.00642 & -0.00301 \bigstrut\\
    \hline
    (450,550)   & 0.0344 & -0.744 & 0.322 & -0.0014 & 0.00167 & -0.000245 \bigstrut\\
    \hline
    (550,650)   & 0.0168 & -0.513 & 0.180  & -0.000446 & -0.000013 & 0.000303 \bigstrut\\
    \hline
    (650,750)   & 0.00649 & -0.376 & 0.114 & 0.0000953 & -0.000682 & 0.000412 \bigstrut\\
    \hline
    (750,850)   & 0.00178 & -0.281 & 0.0768 & 0.000402 & -0.000921 & 0.000397 \bigstrut\\
    \hline
    (850,1000)   & 0.000242 & -0.198 & 0.0501 & 0.000526 & -0.000897 & 0.000323 \bigstrut\\
    \hline
    \end{tabular}%
  \caption{The SM cross sections and the parameterization of the $\kappa_3, \kappa_4$ contributions in the di-Higgs invariant mass bins of $e^-e^+\to \nu \bar{\nu} h h$ at ILC 1 TeV. The ILC beam is polarized as $P(e^-,e^+)=(-0.8, 0.2)$. Note, the $C_{3a}$ and $C_{4b}$ values are independent of the beam polarization.}
  \label{tab:vvhh1000}%
\end{table}%

\end{document}